\documentclass[]{raa}            
\usepackage{graphicx,times}
\usepackage{natbib}

\providecommand{\bm}[1]{\mbox{\boldmath$#1$\unboldmath}}
\def\kms{$\rm{km~s}^{-1}$}
\def\ergs{$\rm{erg~s}^{-1}$}

\begin{document}

   \title{A very bright $i=16.44$ quasar in the `redshift desert' discovered by LAMOST 
}

 \volnopage{ {\bf 2010} Vol.\ {\bf X} No. {\bf XX}, 000--000}
   \setcounter{page}{1}

   \author{Xue-Bing Wu\inst{1}, Zhaoyu Chen\inst{1}, Zhendong Jia\inst{1}, Wenwen Zuo
      \inst{1}
   \and Yongheng Zhao\inst{2}, Ali Luo\inst{2}, Zhongrui Bai\inst{2}, Jianjun Chen\inst{2}, Haotong Zhang\inst{2}, Hongliang Yan\inst{2}, Juanjuan Ren\inst{2}, Shiwei Sun\inst{2}, Hong Wu\inst{2}
   \and Yong Zhang \inst{3}, Yeping Li \inst{3}, Qishuai Lu \inst{3}, You Wang \inst{3}, Jijun Ni \inst{3}, Hai Wang \inst{3}, Xu Kong\inst{4}, Shiyin Shen\inst{5}
   }

   \institute{ Department of Astronomy, Peking University, Beijing 100871, 
China; {\it wuxb@bac.pku.edu.cn}\\
        \and
             National Astronomical Observatories, Chinese Academy of Sciences,
             Beijing 100012, China\\
\and 
National Institute of Astronomical Optics \& Technology, Chinese Academy of
Science, Nanjing 210042, China
\and 
Center for Astrophysis, University of Science \& Technology of China, Hefei 230026, China
\and
Shanghai Astronomical Observatory,  Chinese Academy of Sciences, Shanghai 200030, China
\\
\vs \no
   {\small Received [year] [month] [day]; accepted [year] [month] [day] }
}

\abstract{The redshift range from 2.2 to 3, is known as the 'redshift desert' 
of quasars because quasars with redshift in this range have similar 
optical colors as normal stars and are thus difficult to be found in optical
sky surveys. A quasar candidate, SDSS J085543.40-001517.7, which was 
selected by a recently proposed criterion involving near-IR $Y-K$ and 
optical $g-z$ colors, was identified spectroscopically as a new quasar with 
redshift of 2.427 by the LAMOST commissioning observation in December 2009 
and confirmed by the observation made with the NAOC/Xinglong 2.16m telescope 
in March 2010. This quasar
was not targeted in the SDSS spectroscopic survey because it locates in the stellar
locus of the optical color-color diagrams, while it is clearly separated from
stars in the $Y-K$ vs. $g-z$ diagram. Comparing with other SDSS quasars we 
found this new quasar with $i$ magnitude of 16.44 is apparently the brightest 
one in the redshift range from 2.3 to 2.7. From the spectral properties we derived
its central black hole mass as  $(1.4\sim3.9) \times 10^{10} M_\odot$
and the bolometric luminosity as $3.7\times 10^{48}$ \ergs, which 
indicates that this new quasar is intrinsically very bright and belongs to
the most luminous quasars in the universe. Our identification supports that 
quasars in the redshift desert can be found by the quasar selection
criterion involving the near-IR colors. More missing quasars are expected 
to be recovered by the future LAMOST spectroscopic surveys, which is 
important to the study of the cosmological evolution of quasars at  
redshift higher than 2.2.
}

   \authorrunning{X.-B. Wu et al. }            
   \titlerunning{A very bright quasar in the redshift desert discovered by LAMOST}  
   \maketitle


%
%
\section{Introduction}           
\label{sect:intro}

After the discovery of first quasar (Schmidt 1963), many quasar surveys have
been carried out in optical band and the number of quasars has increased
steadily in the past four decades (Richards et al. 2009). Especially, 
a large number of quasars have 
been discovered in two recent spectroscopic surveys, namely, the Two-Degree 
Fields (2DF) survey (Boyle et al. 2000) and Sloan Digital Sky Survey (SDSS)
(York et al. 2000). 2DF has discovered more than 20,000 quasars (Croom et al. 
2004), and SDSS has identified more
than 100,000 quasars (Schneider et al. 2010; Abazajian et al. 2009).  
2DF mainly selected lower redshift ($z<2.2$) quasars with UV-excess 
(Smith et al. 2005), while SDSS adopted
a multi-band optical color selection method for quasars mainly by
excluding the point sources in the stellar locus of color-color 
diagrams (Richards et al. 2002).  Some 
dedicated methods were also proposed for finding high redshift quasars 
(Fan et al. 2001a,b; Richards et al. 2002).
However, the efficiency
of identifying quasars with redshift between 2.2 and 3 is obviously low in SDSS
(Schneider et al. 2010), because quasars with such redshift usually have 
similar optical 
colors as stars and are thus mostly excluded by the SDSS quasar candidate 
selection algorithm. Therefore, the redshift range
from 2.2 to 3 is often regarded as the `redshift desert' of quasars because
of the difficulty in identifying quasars within this redshift range. 

In addition, the low efficiency of finding quasars with redshift between 2.2 and 3  
has led to the obvious
incompleteness of quasar sample in this redshift range and serious problems in 
constructing the luminosity function for quasars. More importantly, many studies 
have shown that the quasar activity actually peaks at redshift range $2<z<3$ (see
Richards et al. 2006; Jiang et al. 2006). Therefore, recovering
the missing quasars with redshift between 2.2 and 3 has become an important 
task in the quasar study.

Although quasars in the redshift desert have similar optical colors as stars, 
they are usually more luminous than normal stars in the infrared
K-band because the fluxes of normal stars decreases rapidly in the near-IR
bands while quasar SEDs are relatively flat (Warren et al. 2000). 
An important way of finding these missing quasars 
has been suggested by using the infrared K-band excess based on the UKIRT
(UK Infrared Telescope) Infrared Deep Sky Survey (UKIDSS) (Warren et al. 2000;
Hewett et al. 2006; Maddox et al. 2008). Combining the UKIDSS $YJHK$ and SDSS
$ugriz$ magnitudes, some criteria to separate quasars and stars have been
proposed previously. Maddox et al. (2008) suggested a selection criterion
of $z<4$ quasar candidates in the $g-J$ vs. $J-K$ diagram. Chiu et al. (2007)
investigated the different color-color diagrams in optical and near-IR bands
with a sample of 2837 SDSS-UKIDSS quasars, and found that the $g-r$ vs. 
$u-g$ diagram and the $H-K$ vs. $J-H$ diagram are more effective in 
separating quasars and stars
than other diagrams. They also proposed to use the $Y-K$ vs. $u-z$
diagram to select low redshift ($z<3$) quasars. Recently, based on a 
SDSS-UKIDSS sample of 8498 quasars, Wu \& Jia (2010) 
proposed to use the $Y-K$ vs. $g-z$ diagram to select $z<4$ quasars and use 
the $J-K$ vs.$i-Y$ diagram to select $z<5$ quasars. Although with these 
two criteria we can recover 8447 of 8498 SDSS-UKIDSS quasars  
(with a percentage of 99.4\%), we still need to demonstrate whether we can
efficiently discover new quasars, especially those in the redshift desert,  
by applying our criteria to select quasar candidates in the SDSS 
spectroscopically surveyed area. 

The Large Sky Area Multi-Object Fibre Spectroscopic Telescope (LAMOST) is
a 4-meter reflecting Schmidt telescope with 20 square degree field
of view (FOV) and 4000 fibers in the focal plane (Su et al. 1998), located
in the NAOC/Xinglong station. After finishing its main 
construction in 2008, LAMOST has entered the commissioning phase since 2009.
Some test observations have been done in the winter of 2009. Although
LAMOST has not reach its full capability in the commissioning phase, these
observations already led to the discovery of some new quasars, including
a bright quasar with redshift of 2.427, which is the first quasar 
 in the `redshift desert' discovered
by LAMOST.

   \begin{figure}
   \centering
   \includegraphics[width=11.0cm, angle=0]{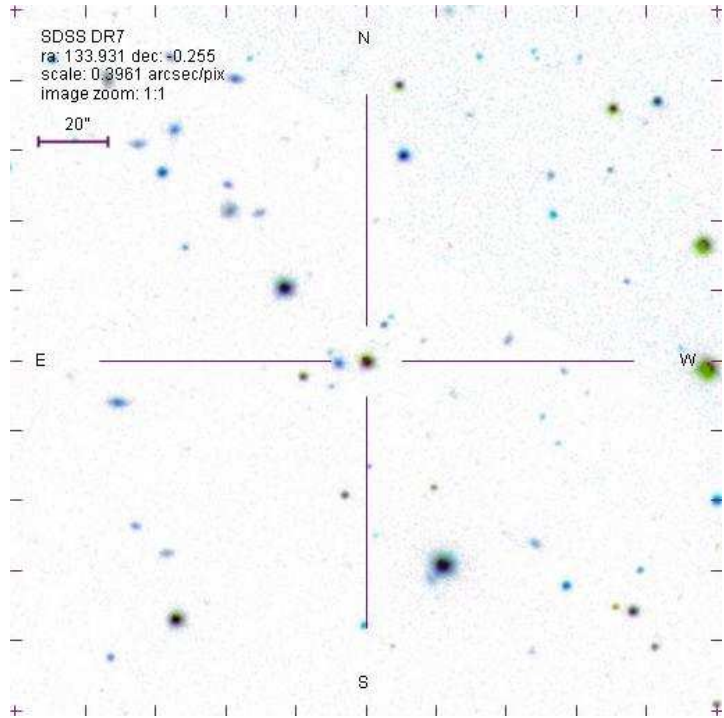}
   \caption{The finding chart of SDSS J085543.40-001517.7. The size is 200"$\times$200". } 
   \label{Fig1}
   \end{figure}

   \begin{figure}
   \centering
   \includegraphics[width=14.0cm, angle=0]{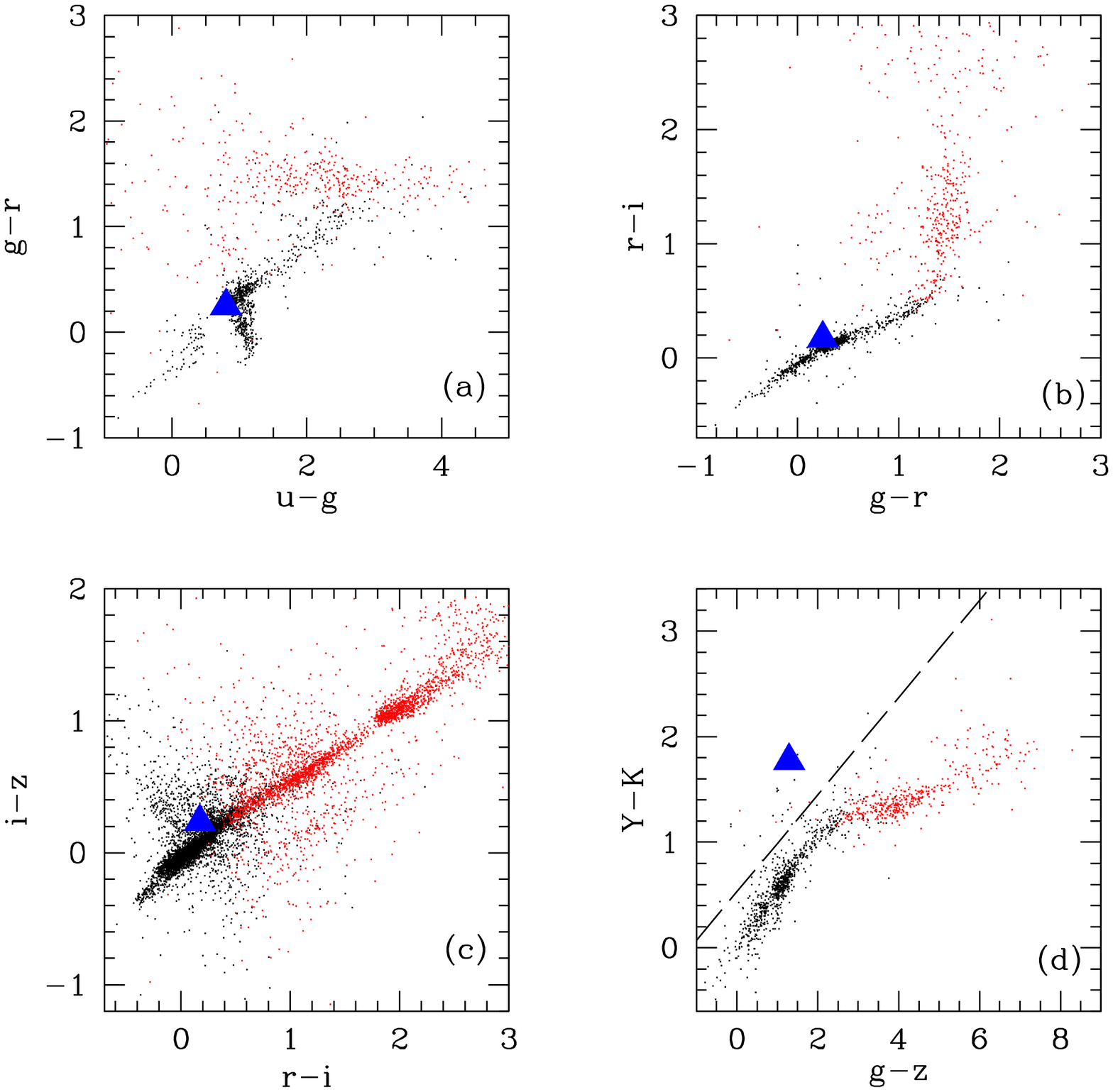}
   \caption{The location of SDSS J085543.40-001517.7 (blue triangle) in three optical color-color diagrams (a,b,c) and
the $Y-K$ vs. $g-z$ diagram (d), in comparing with the 8996 SDSS-UKIDSS stars. Black and red dots represent
the normal and later type stars, respectively. Dashed line shows the $z<4$ quasar selection criterion
proposed by Wu \& Jia (2010).} 
   \label{Fig2}
   \end{figure}

\section{Target selection and Observation}
\label{sect:Obs}

In order to test whether our newly proposed quasar selection criterion in the 
$Y-K$ vs. $g-z$ diagram is efficient in identifying quasars, we selected
many candidates in several sky fields overlapped between UKIDSS and SDSS
survey area with RA from 0 to 9 hours for the LAMOST commissioning observations 
in the winter of 2009. Although LAMOST has met many problems during these
commissioning observations, such as the low accuracy of fiber positioning 
and poor dom seeing condition,  we were still able to identify some new quasars
including one reported here.

SDSS J085543.40-001517.7 is a relatively bright source among our quasar candidates.
After the correction of Galactic extinction using the map of Schlegel et al. (1998), 
Its SDSS $ugriz$ magnitudes (in AB system) are 17.67, 16.87, 16.62, 16.44, 16.20, 
respectively and its UKIDSS $YJHK$ magnitudes (in Vega system) are 15.61,
15.24, 14.60, 13.84, respectively. The offset between its SDSS ad UKIDSS positions
is 0.05$''$. Fig. 1 shows its SDSS finding chart
(obtained from http://cas.sdss.org/dr7/en/tools/chart/chart.asp). Obviously 
SDSS J085543.40-001517.7 is a bright point source, surrounding by several other 
fainter sources with the offsets from 8$''$ to 20$''$. In Fig. 2 we show the location of
this source in 3 optical color-color diagrams and the $Y-K$ vs. $g-z$ diagram, in 
comparison with the 8996 SDSS-UKIDSS stars (Wu \& Jia 2010). 
Note that in the $Y-K$ vs. $g-z$ diagram the magnitude of $g$ and $z$ have been
converted to the magnitudes in Vega system by using the scalings (Hewett et al. 2006): 
$g=g(AB)+0.103$ and $z=z(AB)-0.533$. It is clear that SDSS J085543.40-001517.7 
locates in the stellar locus in three optical color-color diagrams,
but is well separated from stars in the  $Y-K$ vs. $g-z$ diagram and meets the selection
criterion, $Y-K>0.46*(g-z)+0.53$, proposed by Wu \& Jia (2010). This also explains why
this source was not selected as a quasar candidate in SDSS, although it is bright enough.
 
The spectroscopy of SDSS J085543.40-001517.7 was obtained by LAMOST during the 
commissioning observations on December 18, 2009, with the spectral resolution of R$\sim$1000
and the exposure time of 30 minutes. The spectrum was processed using a preliminary version of
LAMOST spectral pipeline. In the left panel of Fig. 3 we show the LAMOST spectrum of 
SDSS J085543.40-001517.7 (some sky light emissions were not well subtracted). From the spectrum we can 
clearly observe at least four strong emission lines, namely, Ly$\alpha$ $\lambda 1216$, 
Si IV $\lambda 1400$, C~IV $\lambda 1549$ and C III] $\lambda 1909$. With these four  
lines we derived an average redshift of $z=2.427$ for this new quasar. Three weak emission lines,
N V $\lambda 1240$, O I $\lambda 1304$ and C II $\lambda 1335$, can be also seen between
 Ly$\alpha$ and Si IV lines. The complicated feature around 5900$\AA$ is due to the 
problem in combining the LAMOST blue and red spectra, which overlap with each other
from 5700$\AA$ to 6000$\AA$.
In this figure we also compare the LAMOST spectrum with the scaled SDSS composite quasar spectrum
(Vanden Berk et al. 2001).   
It is clearly that both match well with each other, except in the red end. 

On March 9, 2010, we also used the NAOC/Xinglong 2.16m telescope to do spectroscopy 
of this new quasar.
Because the seeing condition was bad (4$''$-5$''$), we took two 40-minute exposures on this quasar
and obtained the median spectrum, which is shown in the right panel of Fig. 3 in comparison with
the scaled  SDSS composite quasar spectrum. Although its signal to noise ratio is 
lower than the LAMOST spectrum,  four strong emission
lines can still be clearly observed. Moreover, its  continuum 
shape matches the SDSS composite quasar spectrum better than the LAMOST spectrum, especially in the
red end.

   \begin{figure}
   \centering
   \includegraphics[width=7.0cm, angle=0]{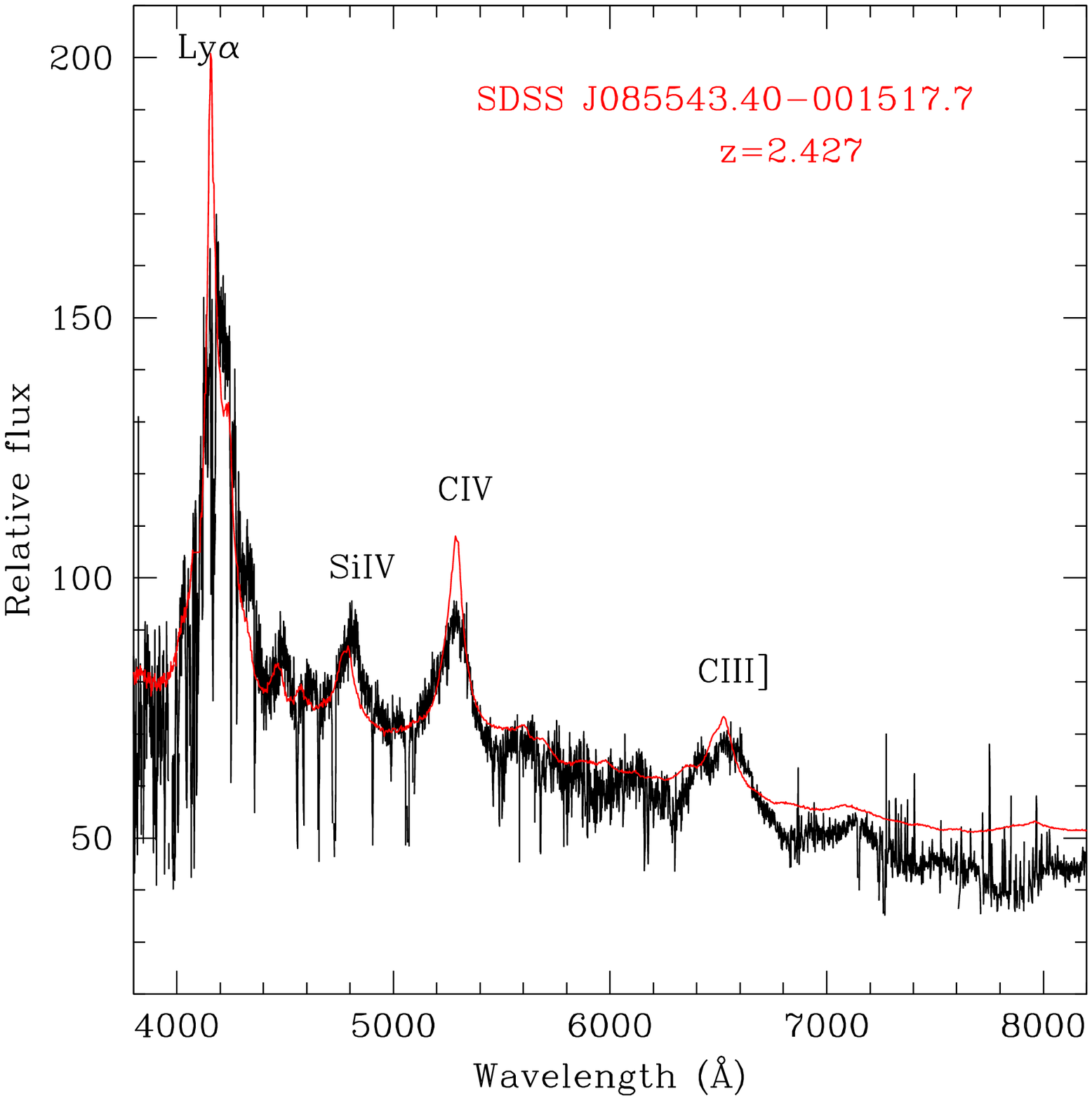}
\includegraphics[width=7.0cm, angle=0]{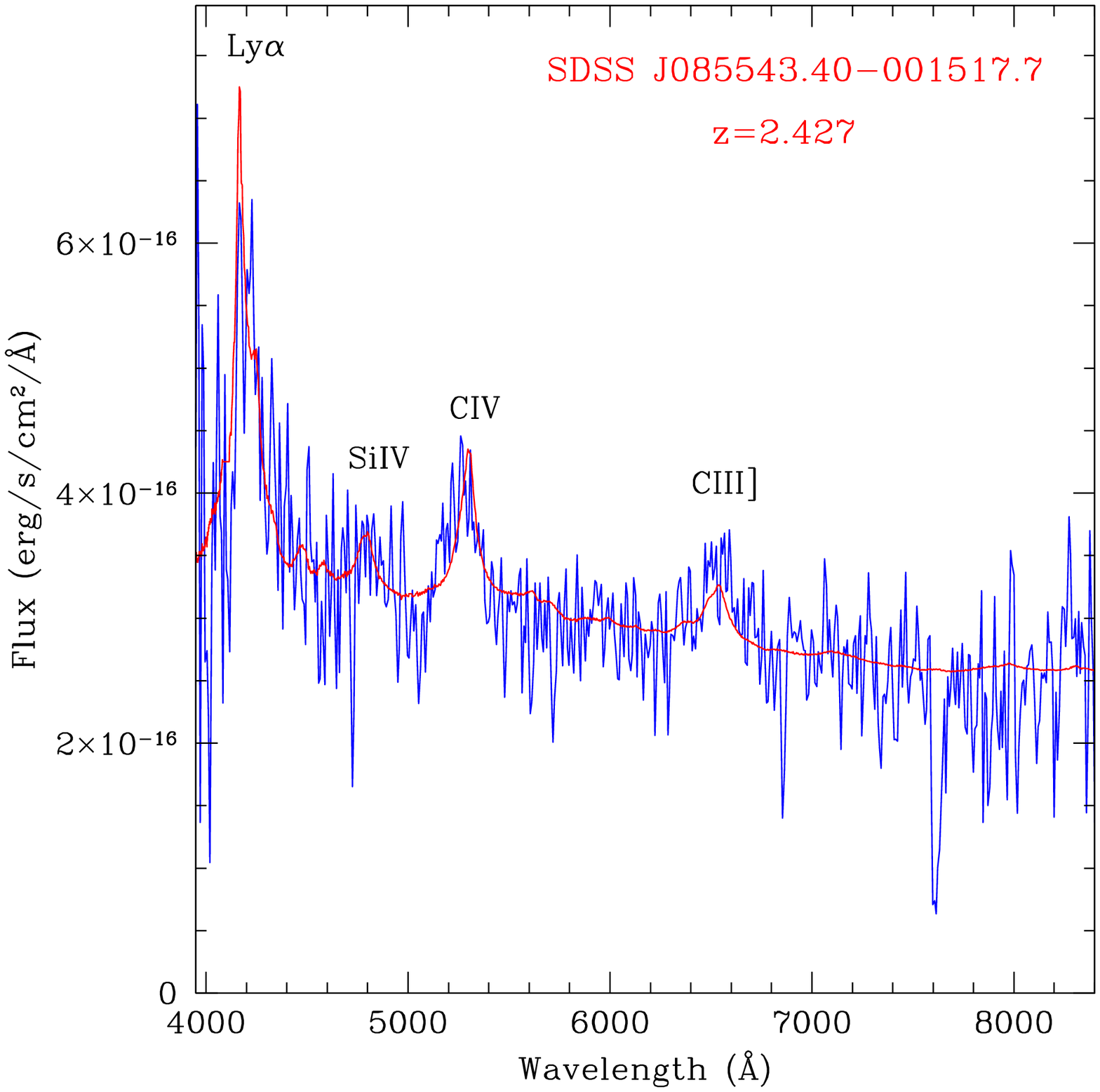}
   \caption{Left panel: The LAMOST spectrum of SDSS J085543.40-001517.7. Right panel: The spectrum of SDSS J085543.40-001517.7 taken by the NAOC/Xinglong 2.16m telescope. 
The scaled SDSS composite quasar spectrum (in red color) is shown for comparison.} 
   \label{Fig3}
   \end{figure}

 \begin{figure}
   \centering
   \includegraphics[width=14.0cm, angle=0]{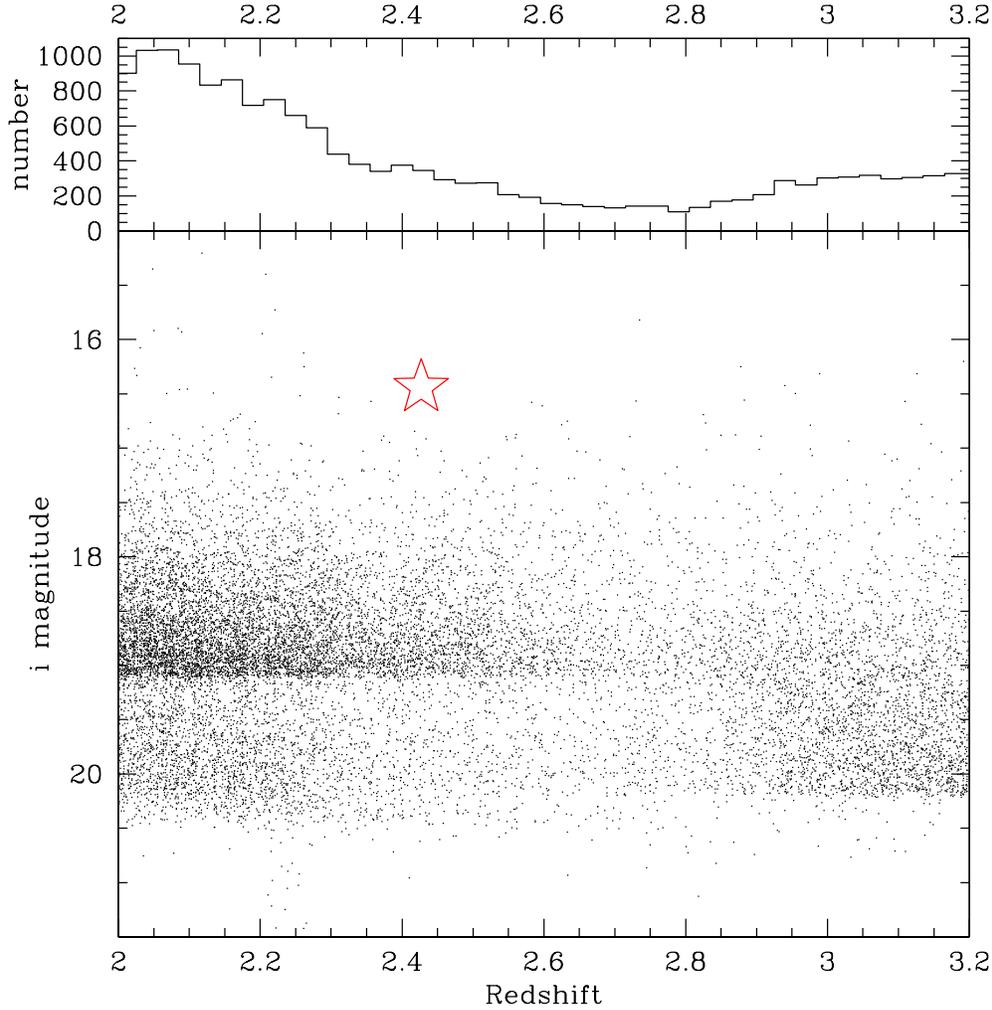}
   \caption{The location of SDSS J085543.40-001517.7 (red star) in the magnitude-redshift diagram in
comparison with the SDSS DR7 quasars in redshift range from 2 to 3.2. The redshift distribution
of SDSS quasars is also shown in the upper panel, in which the redshift desert ( with redshift
 from 2.2 to 3) is clearly presented.
 } 
   \label{Fig4}
   \end{figure}

\section{Properties of SDSS J085543.40-001517.7}
With the $i$ magnitude of 16.44 and redshift of 2.427, SDSS J085543.40-001517.7 is undoubtedly
a very bright quasar. We compared it with other SDSS quasars in the redshift range from
2 to 3.2 and found the new quasar is indeed very bright.  In Fig. 4 we show the location of 
the new quasar in the magnitude-redshift diagram in comparison with other SDSS quasars,
as well as the histogram of the redshift distribution of SDSS quasars. The redshift
distribution clearly shows the presence of `redshift desert' in the redshift range
from 2.2 to 3. The new quasar is  apparently
the brightest one in the redshift range from 2.3 to 2.7. Its absolute $i$ magnitude
is -30.0 if the cosmological parameters $\rm{H_0}$=70 \kms Mpc$^{-1}$, 
$\Omega_M=0.3$ and $\Omega_\Lambda=0.7$ 
are adopted. Clearly this quasar belongs to the most luminous quasars in the universe.

We also searched the counterparts of  SDSS J085543.40-001517.7 in other wavelength bands.
From the VLA/FIRST radio catalog (White et al. 1997) we did not find any radio counterpart 
within 20$''$ from its
SDSS position. The closest radio source is 121.5$''$ far away. Therefore, this quasar is a 
radio-quiet one, which is also another reason why it is missed by the SDSS spectroscopy. 
We also searched the ROSAT X-ray source catalog (Voges et al. 1999) 
and did not find any
counterpart within 1'. The closest X-ray source is 23$'$ away. From GALEX catalog 
(Morrissey et al, 2007) we failed to 
find any ultraviolet counterpart within 5$''$. One GALEX source is 27$''$ away (in the 
south-western direction) from the optical position of 
SDSS J085543.40-001517.7, and is clearly the  counterpart of another fainter extended
source in the SDSS image. Therefore, we believe that 
SDSS J085543.40-001517.7 is faint in radio, UV and X-ray bands, although it is 
very luminous in optical and near-IR bands. 

From the spectral properties we can estimate the black hole mass and bolometric luminosity
of this new quasar. After doing the redshift correction, Galactic extinction correction 
using the reddening map of Schelegal et al. (1998), continuum fitting and Fe II
subtraction using the template from Vestergaard \& Wilkes (2001), we measured the C~IV line 
width (FWHM, the Full Width at Half Maximum) and the rest frame 1350$\AA$ continuum flux
from the spectrum. 
Because we did not make the absolute flux calibration of the LAMOST spectrum of SDSS J085543.40-001517.7,
we used the ultraviolet continuum window 1320$\rm \AA$ -- 1330$\rm \AA$ 
 to calibrate the LAMOST spectrum with the spectrum 
taken by the 2.16m telescope. The C~IV FWHM values measured from the LAMOST and 2.16m spectra are
8520\kms and 11040\kms, respectively. Due to the lower signal to noise ratio of the 2.16m
spectrum (see the right panel of Fig. 3), we took the C~IV FWHM value from the LAMOST spectrum in the following calculation.
The black hole mass estimation was done with two similar formula proposed by Kong et al. (2006)
and Vestergaard \& Peterson (2006), both involving the C~IV line width and $1350\AA$ continuum
luminosity. The first one gives $M_{BH}=1.4\times 10^{10}M_\odot$ and the latter one
gives  $M_{BH}=3.9\times 10^{10}M_\odot$. Using a scaling between 1350$\AA$ luminosity and bolometric 
luminosity, $L_{bol}=4.62L_{1350}$, given by Vestergaard (2004) based on the SED of radio-quiet 
quasars (Elvis et al. 1994), 
we estimated the bolometric
luminosity of this new quasar as $3.7\times 10^{48}$\ergs, which is about $(0.5\sim1.4)$ times of
the Eddington luminosity if the above estimated black hole mass is adopted. Obviously, 
this quasar is
intrinsically very bright, and is accreting matters with the accretion rate around
the Eddington limit.  

\section{Discussion}
\label{sect:discussion}

Quasars with redshifts in the range from 2.2 to 3  are very important for studying 
their cosmological evolution, and the relation between quasar activity and star 
formation activity which peaks at redshift between 1 and 2 (Madau et al. 1998). 
However, because these quasars have similar optical colors as normal stars, it is 
very difficult for find them in previous quasar
surveys. The low efficiency of finding quasars in the `redshift desert' has led to the obvious
incompleteness of quasar sample in this redshift range and serious problems in 
constructing the luminosity function for quasars around the redshift peak 
(between 2 and 3) of quasar activity (Richards et al. 2006; Jiang et al. 2006). 

In this paper we have presented a case study to find a very bright new quasar in the 
redshift desert by the LAMOST commissioning observation. The spectroscopic 
identification of an $i=16.44$ source, SDSS J085543.40-001517.7, as a $z=2.427$ 
quasar gives us confidence to discover more missing quasars in the future 
LAMOST quasar survey.  
This discovery also supports the idea that by combining the UKIDSS near-IR colors with the
SDSS optical colors we are able to efficiently recover these missing quasars. 
In the winter of 2009, LAMOST has made test
observations on several sky fields and we are now searching for more quasars from
the spectra taken in these fields. The discovery of more new quasars in these
fields will be reported in the future works. We hope that in the next a few months great
progress will be made in improving the capability of LAMOST spectroscopy and the
spectral processing pipeline. As long as LAMOST 
can reach its designed capability after the commissioning phase, we expect to find
several hundred-thousands of quasars in the LAMOST quasar survey. This will form
the largest quasar sample in the world and play a leading role in the quasar study
of the next decade.

\normalem
\begin{acknowledgements}
This work was supported by the National 
Natural Science Foundation of China  (10525313), the
National Key Basic Research Science Foundation of China (2007CB815405),
and the Open Project Program of the Key Laboratory of Optical Astronomy, 
NAOC, CAS. The Large Sky Area Multi-Object Fiber Spectroscopic Telescope (LAMOST)
is a National Major Scientific Project built by the Chinese Academy of
Sciences. Funding for the project has been provided by the National
Development and Reform Commission. The LAMOST is operated and managed
by the National Astronomical Observatories, Chinese Academy of Sciences.
We acknowledge the use of LAMOST and the NAOC/Xinglong 
2.16m telescope, as well as
the archive data from SDSS, UKIDSS, FIRST, ROSAT and GALEX. We thank Marianne 
Vestergaard
for kindly providing us the Fe II template, which was used in this work.

\end{acknowledgements}


\end{document}